\documentclass[preprint,showpacs,preptintnumbers]{revtex4}
\usepackage{amssymb}
\usepackage{graphicx}
\usepackage{dcolumn}
\usepackage{times}
\usepackage{textcomp}
\usepackage{bm}
\begin{document}

\title{Predicted formation of superconducting platinum-hydride crystals under pressure in the presence of molecular hydrogen}

\author{Duck Young Kim$^{1}$}
\author{Ralph H.\ Scheicher$^{2}$}
\author{Chris J.\ Pickard$^{3}$}
\author{R.\ J.\ Needs$^{1}$}\email{rn11@cam.ac.uk}
\author{R.\ Ahuja$^{2,4}$}\email{rajeev.ahuja@fysik.uu.se}

\affiliation{$^{1}$ Theory of Condensed Matter Group, Cavendish Laboratory, J J Thomson Avenue, Cambridge CB3 0HE, United Kingdom \\
$^{2}$ Condensed Matter Theory Group, Department of Physics and Astronomy, Uppsala University, P.O.\ Box 516, SE-751 20, Uppsala, Sweden \\
$^{3}$ Department of Physics and Astronomy, University College London, Gower St, London WC1E 6BT, United Kingdom \\
$^{4}$ Department of Materials and Engineering, Applied Materials Physics, Royal Institute of Technology (KTH), SE-100 44 Stockholm, Sweden
}

\begin{abstract}
  Noble metals adopt close-packed structures at ambient pressure and
  rarely undergo structural transformation at high pressures.
  Platinum (Pt), in particular, is normally considered to be
  unreactive and is therefore not expected to form hydrides under
  pressure.  We predict that platinum hydride (PtH) has a lower
  enthalpy than its constituents solid Pt and molecular hydrogen at
  pressures above 21.5 GPa.  We have calculated structural phase
  transitions from tetragonal to hexagonal close-packed or
  face-centered cubic (fcc) PtH between 70 and 80 GPa.  Linear
  response calculations indicate that PtH is a superconductor at these
  pressures with a critical temperature of about 10--25 K. These
  findings help to shed light on recent observations of
  pressure-induced metallization and superconductivity in
  hydrogen-rich materials. We show that formation of fcc metal
  hydrides under pressure is common among noble metal hydrides and
  examine the possibility of superconductivity in these materials.
\end{abstract}

\pacs{63.20.kd, 74.25.Jb, 74.10.+v, 74.62.Fj}

\maketitle



Pt maintains the face-centered cubic crystal structure up to at least
304 GPa \cite{Rajeev:1994}, which covers most of the range currently
accessible in static diamond anvil cell (DAC) experiments, and makes
it suitable for use as a pressure standard
\cite{Holmes:1989,Tonkov:2005}. Pt is frequently used to form
electrodes (which allow measurements of electrical resistance, the
concentration of electrical carriers and the Hall mobility) and as a
heat coupler.  As a noble metal, Pt is fairly unreactive at ambient
conditions, and this has generally been assumed also to hold at high
pressures.  However, chemical reactions of Pt with other materials
present in the sample chamber can lead to the unexpected formation of
new compounds as, for example, described by Gregoryanz \textit{et al.}
\cite{Gregoryanz:2004} who observed formation of Pt nitride when
heating nitrogen with a piece of platinum.


The hydrogen-dense material silane (SiH$_{4}$) forms molecular
crystals at low pressures.  Two experimental studies of the
metallization of SiH$_{4}$ under pressure have recently appeared
\cite{Mao:SiH4,Eremets:SiH4}. The first study \cite{Mao:SiH4} provided
optical evidence for metallization of SiH$_{4}$ above 60 GPa.
Remarkably, the second study by Eremets \textit{et al.}\
\cite{Eremets:SiH4} reported superconductivity of SiH$_{4}$ in the
pressure range 60--150 GPa with a superconducting transition
temperature as high as 17 K. These experimental findings have
motivated a number of theoretical investigations of compressed
SiH$_{4}$
\cite{Kim:PNAS2008,PhysRevLett.101.077002,PhysRevLett.102.087005}
suggesting that metallization and/or superconductivity can be achieved
under pressure, but a clear explanation of the observations has not
yet been achieved. While the above two experimental studies
\cite{Mao:SiH4,Eremets:SiH4} reported metallization of SiH$_{4}$ above
60 GPa, whereas some other experiments claimed decomposition,
amorphization, and the formation other metal-hydrogen alloys besides
silane \cite{Narayana:2008, Degtyareva}. This complexity itself is
associated with hydrogen as being extremely light and mobile, as well
as extraordinarily reactive. The properties of hydrogen are becoming
drastically enhanced under high pressure and temperature, and make the
interpretation of experimental data for hydrides \cite{Baranowski1,
  Baranowski2, Antonov1984} rather inconclusive.


Degtyareva \textit{et al.}\ \cite{Degtyareva} have speculated that the
formation of PtH under high pressures might explain some of the
observations reported in Ref.\ \citenum{Eremets:SiH4}, and they
provided indirect evidence to support the hypothesis of PtH formation.
Moreover, it appears likely that the silane in the experiments of
Eremets \textit{et al.}\ \cite{Eremets:SiH4} partially decomposed,
leading to release of hydrogen.  Indeed SiH$_{4}$ is a metastable
compound at low pressures which is thermodynamically unstable to
decomposition into H$_2$ and Si \cite{Degtyareva}, so that release of
hydrogen might occur when the compression is sufficient to break bonds
in the SiH$_{4}$ molecules.  Released hydrogen may then have reacted
with the Pt electrodes to form PtH as the pressure was further
increased.

Confirming the formation of superconducting PtH at high pressures
would change the interpretation of the data of Eremets \textit{et
  al.}\ \cite{Eremets:SiH4}, but would in no way diminish the
importance of the experiments.  In the new interpretation these
experiments will have led to the discoveries of superconducting PtH
under pressure and a new insulating phase of SiH$_{4}$, and to the
understanding that compressed SiH$_{4}$ is prone to decomposition.
Indeed, Degtyareva \textit{et al.}\ \cite{Degtyareva} have suggested
that hydrogen release from SiH$_{4}$ at high pressures could be useful
in synthesising metal hydrides.


The prevalence of DAC measurements and the steadily growing interest
in high-pressure experimental studies of metal hydrides raises the
possibility of decomposition and hydrogen release in such experiments.
Inadvertent and undetected formation of noble metal hydrides could
lead to erroneous pressure estimates and other effects which could
substantially alter the interpretation of the data.  A careful
investigation of the possibility of pressure-induced reactions between
hydrogen and noble metals such as Pt is therefore very important and
timely.


We have used \textit{ab initio} random structure searching (AIRSS) to
predict crystal structures of PtH under pressure
\cite{Pickard_silane}.  This approach has been used to predict the
structures of materials such as hydrogen \cite{Pickard_hydrogen},
metal hydrides \cite{Pickard_aluminum_hydride}, transition metals
\cite{Pickard_iron}, and molecular crystals
\cite{Pickard_ammonia,Fortes_ammonia_monohydrate,Srepusharawoot}.  In
this study we have used the simplest version of the AIRSS technique in
which an ensemble of structures is prepared by generating unit cells
of random shapes with reasonable volumes into which the atoms are
placed at random positions.  The cell shapes and atomic positions are
then relaxed to an enthalpy minimum at fixed pressure.  We studied
simulation cells containing 1, 2, 4, 6, and 8 formula units of PtH. We
found three particularly low-enthalpy metallic structures of PtH (the
corresponding space groups are given in parentheses): fcc
($Fm\bar{3}m$), hcp ($P6_3/mmc$), and tetragonal ($I\bar{4}m2$).

\begin{figure}[ht]
\centering
\includegraphics[width=8cm, angle=0]{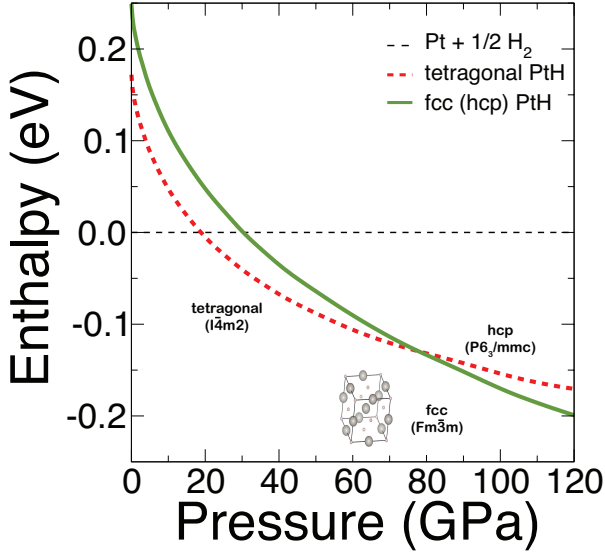}
\caption{(color online). Comparison of the enthalpies of Pt + $\frac{1}{2}$H$_2$ and
  the tetragonal and fcc phases of PtH.}
\label{enthalpy}
\end{figure}

We have assumed that PtH forms in the 1:1 stoichiometry.  The fcc
structure was used for Pt and an hcp molecular form for hydrogen
\cite{Pickard_hydrogen}, in which the H$_2$ molecules are randomly
oriented in space.  The variation of the enthalpies $H$ of the fcc,
hcp, and tetragonal phases of PtH relative to that of Pt +
$\frac{1}{2}$H$_2$ is given in Fig.\ \ref{enthalpy}, which shows that
PtH is more stable at pressures above approximately 20 GPa. We also
calculated the dynamical stability of the PtH structures and evaluated
their electron-phonon coupling (EPC) strengths, $\lambda$, in order to
study the possibility of phonon-mediated superconductivity in PtH.
Since the zero-point motion of the hydrogen nuclei is substantial, we
have included effects from the zero-point enthalpy (ZPE) calculated
within the quasiharmonic approximation. For details of the
computational methods we refer the reader to the supplementary
material \cite{supplmat}. It should be noted that there are
alternative ways of calculating the superconducting transition
temperature $T_c$. In particular, there has recently been considerable
progress within the framework of density functional theory in
developing truly \emph{ab initio} quantitative predictions of $T_c$ in
the absence of adjustable parameters, which combine electron-phonon
coupling and Coulomb interactions
\cite{EKUGross:2005a,EKUGross:2005b,EKUGross:2005c,EKUGross:2006,EKUGross:2007a,EKUGross:2007b}.

Figure \ref{enthalpy} shows that it is thermodynamically favourable
for tetragonal PtH to form from elemental Pt and H$_2$ above 18.7 GPa
when the zero-point energy of the nuclei is neglected (above 21.5 GPa
when the zero-point energy contribution is included; see the
supplementary material \cite{supplmat}).  Experimentally, a tetragonal
PtH was observed at room temperature, 26 GPa on compression and 20 GPa
on decompression \cite{Hirao:PtH}.  We find a transition from
tetragonal to hcp or fcc PtH between 70 and 80 GPa (Fig.\
\ref{enthalpy}).  The fcc and hcp structures differ only in the
stacking of layers, making them virtually degenerate in energy (to
within a few meV per formula unit).  Under pressure, a previous
calculation using a simple model suggested that the site occupied by
hydrogen in $d$-metal mono-hydrides tends to change from the
tetrahedral to the octahedral site and that the hydride transforms
into the hcp or fcc structure \cite{somenkov:1987}, in good agreement
with our findings.  In the present context it is more likely for fcc
PtH to form, since the Pt atoms as the heavier host material are
arranged in that structure prior to exposure to hydrogen.
Experimentally hcp PtH not fcc PtH was observed and we therefore
speculate that kinetics might favor a sequence of structural
transformation from tetragonal PtH to hcp PtH.


Having established the likely formation of PtH from Pt and H$_2$ in
the pressure range relevant for the experimental observations reported
in Ref.\ \citenum{Eremets:SiH4}, we evaluated the electron-phonon
coupling (EPC) strength in PtH to examine its potential for
superconductivity, in a similar fashion to our studies of metal
tri-hydrides \cite{Kim:2010, Kim:2009}, as summarized in the
supplementary material \cite{supplmat}.  This study is crucial because
the formation of superconducting SiH$_{4}$ was a central claim of
Ref.\ \citenum{Eremets:SiH4}, and perhaps superconducting PtH could
provide an alternative explanation.

\begin{figure}[ht]
\centering
\includegraphics[width=8cm, angle=0]{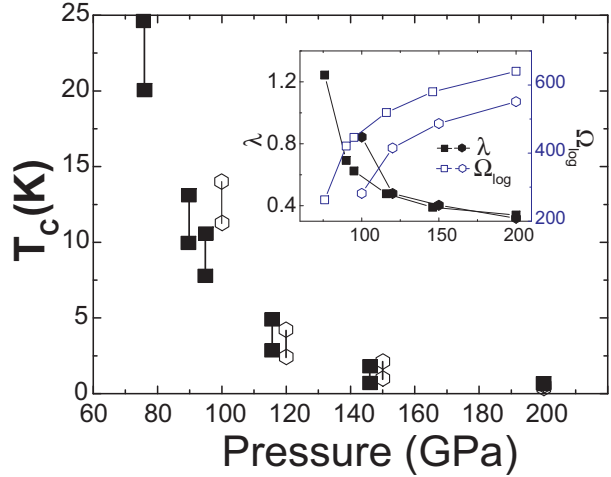}
\caption{(color online). Superconducting transition temperature $T_c$ of PtH as a
  function of pressure. Filled squares and hollow hexagons show values
  of $T_c$ for the fcc and hcp structures, respectively, with the
  higher $T_c$ value at each pressure corresponding to the choice
  $\mu^{*}$=0.10 and the lower $T_c$ value to the choice
  $\mu^{*}$=0.13. The inset displays the evolution with pressure of
  the electron-phonon coupling parameter, $\lambda$, and the
  asymptotic phonon momentum $\Omega_{log}$ (see the supplementary
  material \cite{supplmat}). The fcc phase is dynamically unstable below 77 GPa and the hcp phase is dynamically unstable below 100 GPa.}
\label{Tc}
\end{figure}


Papaconstantopoulos suggested that PtH could be superconducting under
ambient conditions \cite{papa1980}, but subsequent experiments found
no evidence of PtH formation at low pressure
\cite{traverse1981,Hirao:PtH}. Tetragonal PtH becomes more stable than
Pt + $\frac{1}{2}$H$_2$ above 20 GPa.  We calculated the EPC of
tetragonal PtH in the pressure range 20--60 GPa and found no
significant coupling, so it is not likely that superconducting
tetragonal PtH could exist.  Figure \ref{Tc} shows the variation of
the calculated superconducting transition temperature $T_c$ of PtH
with pressure.
In the whole pressure interval of 80--200 GPa we used a fixed value for
the effective Coulomb interaction parameter $\mu^{*}=0.13$, as has
been common procedure \cite{ashcroft:2004,PhysRev.167.331}. However, we
also explored the effects on $T_c$ of changing $\mu^{*}$ to 0.10
\cite{PhysRev.167.331,PhysRevB.12.905}. With this, we have covered the
two extreme choices of $\mu^{*}$, resulting in corresponding high- and
low-estimates of $T_c$ (Fig.\ \ref{Tc}). It is worth noting that the value of $\mu^{*}$
may also be chosen to vary as a function of the electronic density of
states at the Fermi level \cite{Bennemann:1972}.  The highest value of
$T_c$ is found in the vicinity of the onset of dynamical stability of
the structures, as expected for phonon-mediated superconductivity
\cite{Kim:2010, Li:2009, Si:1986}. The fcc structure is dynamically
stable above 77 GPa where it possesses the highest $T_{c}$ of around
25 K.  $T_{c}$ decreases monotonically and quite rapidly as the
pressure rises, as shown in Fig.\ \ref{Tc}, and reaches essentially
zero around 140 GPa.  The hcp structure is dynamically stable above
100 GPa (and is almost degenerate in enthalpy with fcc PtH) and its
maximum $T_{c}$ of around 15 K occurs at 100 GPa.


\begin{figure}[ht]
\centering
\includegraphics[width=14cm, angle=0]{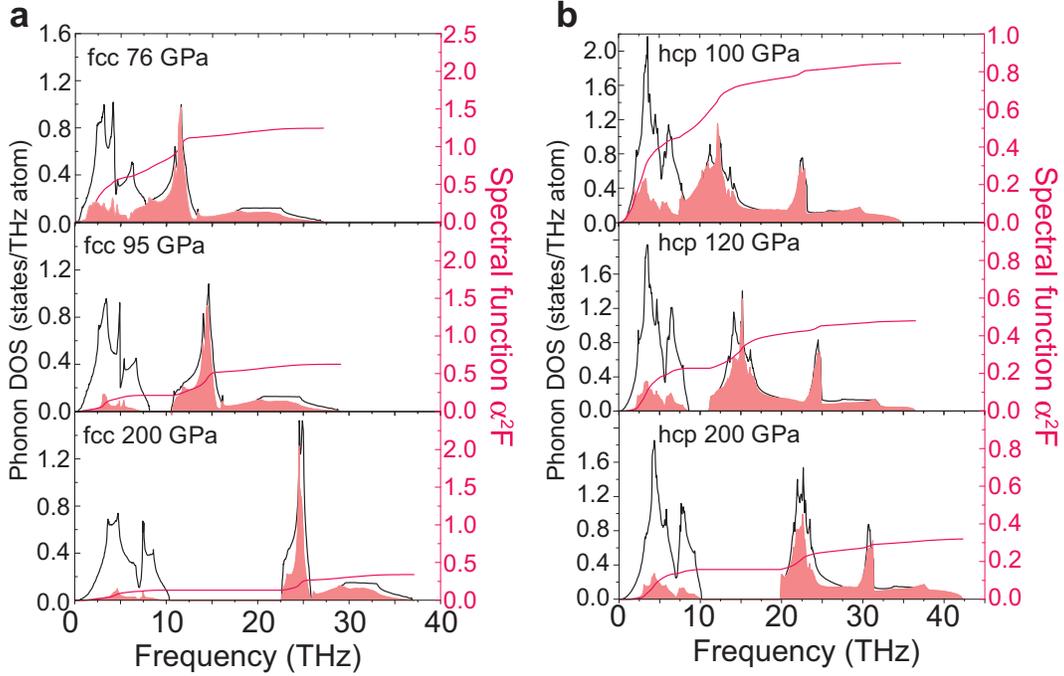}
\caption{(color online). Spectral function $\alpha^{2}F$ (red shaded areas), the
  integral of the spectral function up to frequency $\omega$
  ($\int_0^{\omega} \alpha^{2}F(\omega^{\prime})/\omega^{\prime} \,
  d\omega^{\prime}$) (red line), and the phonon density of states
  (solid black line) as a function of frequency at selected pressures
  for (a) fcc PtH and (b) hcp PtH.}
\label{spectral}
\end{figure}

Figure \ref{spectral} shows the spectral function $\alpha^{2}F$ and
the phonon density of states (PhDOS) at different pressures for both
fcc and hcp PtH.  At the lowest pressure at which the fcc phase is
stable (77 GPa), the Pt and H atoms both contribute to the EPC (Fig.\
\ref{spectral}a).  With increasing pressure, the spectral function
splits into high and low frequency bands as shown in the second and
third panels of Fig.\ \ref{spectral}a, and only hydrogen vibrations
contribute to the EPC in the higher pressure band.  The situation is
similar in the hcp phase (Fig.\ \ref{spectral}b), in the sense that
the respective contributions to the EPC parameter $\lambda$ from Pt
and H are nearly the same as at lower pressures. With increasing
pressure, the integrated $\lambda$ decreases as the phonon frequencies
increase.  This behavior can be understood from Eq.\ 3 of the
supplementary material \cite{supplmat}, where it can be seen that the
spectral function $\alpha^{2}F$ is divided by the phonon frequencies.

Interestingly, Eremets \textit{et al.}\ reported an apparent
divergence in $T_c$ between 80 and 100 GPa \cite{Eremets:SiH4} which
coincides with our calculated transition pressure between fcc and hcp
PtH. These structural transformations are accompanied by softening of
the acoustic branch modes which is a precursor of structural
transformation.  High-pressure experiments on PtH have been reported
earlier \cite{Hirao:PtH}, but since they were performed at room
temperature, no information regarding possible superconductivity could
be obtained.  It would be very interesting to study compressed PtH at
temperatures below 25 K to determine whether it is indeed
superconducting.

At low pressures, fcc-Pt absorbs hydrogen and forms tetragonal PtH,
which is a distortion of the fcc structure.  Under pressure, the
tetragonal PtH phase is restored to its original fcc structure or the
similar close-packed hcp structure, which possess strong EPC and hence
a high superconducting critical temperature.  Such a structural
transformation from a lower-symmetry phase to the fcc structure also
applies to other hydrides formed by metals which are neighbors of Pt
in the periodic table (referred to in the following as the Pt family
of hydrides).  We investigated the metals Rh, Pd, Ag, Ir, and Au,
which all adopt the fcc structure at ambient pressure and have
extremely small or zero superconducting $T_c$,
and determined the onset pressure at which the corresponding fcc
hydrides become dynamically stable.  We find that fcc-RhH (lattice
constant $a$ = 4.02 \AA) and fcc-PdH ($a$ = 4.08 \AA) are stable at
almost ambient pressure, consistent with experimental observations
\cite{RhH,RhH2, fukai, PdH}. Also, our calculations are in good
agreement with the measured lattice constants which were reported to
be 4.01 \AA\ at 3.8 GPa (hydrogen gas pressure) for fcc-RhH \cite{RhH,
  RhH2} and 4.09 \AA\ at ambient pressure for PdH \cite{PdH}.  The
metal hydrides fcc-AgH ($a$ = 4.00 \AA) and fcc-IrH ($a$ = 3.88 \AA)
are dynamically stable at 50 and 80 GPa, respectively.  Dynamical
stability of fcc-AuH ($a$ = 3.80 \AA) is achieved only above 220 GPa.
We emphasize that other hydride structures besides fcc might be stable
at lower pressures.

\begin{figure}[ht]
\centering
\includegraphics[width=9cm, angle=0]{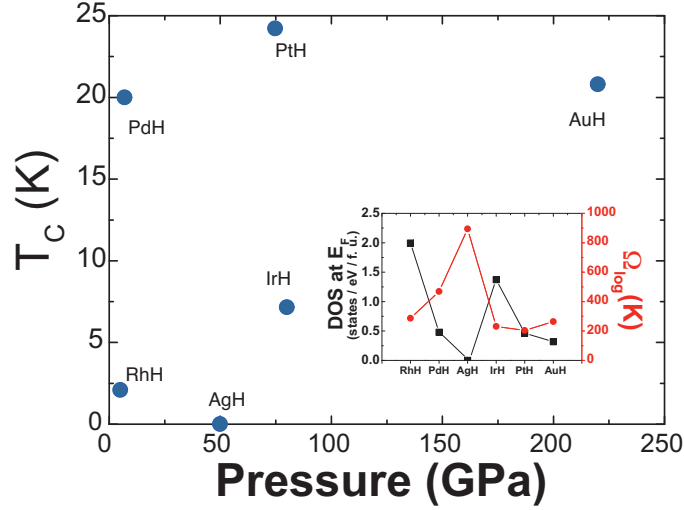}
\caption{(color online). Maximum $T_c$ (calculated at the onset of dynamical stability
  of the fcc phase) for the metal hydrides RhH, PdH, AgH, IrH, PtH,
  and AuH. Note that the corresponding elemental metals in their fcc
  form all possess either extremely small $T_c$ (Rh and Ir) or are not
  superconducting at all (Pd, Ag, Pt, and Au).}
\label{TcofMH}
\end{figure}

Similarly to PtH, the dynamical stability of other fcc-structure
hydrides can be regarded as precursors for superconductivity, while
lower symmetric structures might form at much lower pressures
\cite{Antonov:1982}.  Figure \ref{TcofMH} shows the calculated $T_c$
at the onset of dynamical stability of the fcc structure in the Pt
family of hydrides.  PdH is calculated to possess a $T_c$ of 20 K,
which is compatible with experiments on non-stoichiometric PdH$_x$
which reported a $T_c$ of 9 K \cite{PdH}, although we note that our results do not reproduce the observed isotope dependence of $T_c$ in PdH/PdD \cite{Stritzker:1972}.  We predict IrH to have a
$T_c$ of 7 K at 80 GPa.  Remarkably, the $T_c$ of fcc-AuH is estimated
to be 21 K, however, it takes a very high pressure of 220 GPa to reach
this superconducting state.  As the pressure is increased, the $T_c$
of all these fcc-hydrides will decrease, analogously to the case of
PtH.  We predict that AgH is not a superconductor, because the electronic density of states exhibits a small bandgap at the Fermi level. AgH does however possesses the highest
$\Omega_{log}$, and therefore, if electrons were to be donated in a complex alloy, this could lead to a good superconducting material, such as (PdAg)H$_{x}$, possessing a $T_{c}$ of about 16 K \cite{Buckel:1973}. RhH has a
small $T_{c}$ and it is worth note that no superconductivity of RhH
has been detected at $T_{c} >$ 0.3 K \cite{Antonov:1987}.

In conclusion, we have shown from first principles that formation of
PtH from Pt and H$_2$ can occur under pressure.  Based on the
empirical Allen-Dynes equation (see supplementary material), the
resulting product is predicted to be a superconductor.  Throughout the
studied pressure interval of 80--200 GPa, we used the effective
Coulomb interaction parameter $\mu^{*}= 0.13$, a value which has often
been used \cite{ashcroft:2004}. We did, however, consider the effect
of changing $\mu^{*}$ to 0.10 on the predicted $T_c$.  If our
prediction concerning the formation of superconducting PtH at high
pressures would be experimentally confirmed then this might lead to
new interpretation of the data of Eremets \textit{et al.}\
\cite{Eremets:SiH4}.  That work would mark the discovery of
superconducting PtH under pressure and a new insulating phase of
SiH$_{4}$, and to the understanding that compressed SiH$_{4}$ is prone
to decomposition.  We expect these findings to be relevant to recent
observations of pressure-induced metallization and superconductivity
in hydrogen-rich materials. Indeed, our calculations show that the
formation of superconducting fcc metal hydrides under pressure may be
a common phenomenon among noble metal hydrides. The pressure-induced
formation of PtH and of other noble metal hydrides, along with their
associated superconductivity, represents an important aspect which
should be considered in the interpretation of diamond anvil cell
experiments whenever the corresponding metals and hydrogen are
present, with the latter introduced either directly or entering via
decomposition of hydrides.

The authors acknowledge financial support from Wenner-Gren
Foundations, the Swedish Research Council (VR, Grant No.\
621-2009-3628), Formas, and the Engineering and Physical
Sciences Research Council of the UK. The Swedish National
Infrastructure for Computing (SNIC) and the Uppsala Multidisciplinary
Center for Advanced Computational Science (UPPMAX) provided computing
time.

\newpage

\section*{Supplementary Material}

\begin{center}
  ``Formation and superconductivity of platinum hydride under pressure''\\
  D. Y. Kim, R. H. Scheicher, C. J. Pickard, R. J. Needs, and R. Ahuja
\end{center}

\subsection{Methods}

Our results were obtained using density-functional-theory (DFT)
methods \cite{PhysRev.136.B864,PhysRev.140.A1133} and the generalized
gradient approximation (GGA) of Perdew, Burke and Ernzerhof for the
exchange-correlation functional
\cite{PhysRevB.46.6671,PhysRevLett.77.3865}.  The structure searching
was carried out with the \textsc{CASTEP} plane-wave code \cite{castep}
and ultrasoft pseudopotentials \cite{up}. For the searches we used a
plane-wave basis-set cutoff of 550 eV and a Brillouin-zone integration
grid of spacing 2$\pi \times$ 0.07 \AA$^{-1}$. The structures of
interest were further relaxed for enthalpy comparisons, which were
done with the VASP code \cite{vasp1,vasp2} with the
projector-augmented wave (PAW) approach \cite{PAW}. A cutoff energy of
700 eV was found to produce converged results. The Monkhorst-Pack
scheme \cite{PhysRevB.13.5188} was used to generate the $k$-point grid
meshes of $31\times31\times31$ for Pt and PtH ($\Gamma$-centered in
the case of hcp-PtH). For calculations of solid hydrogen, a large
supercell containing 36 H$_2$ molecules was used; therefore, the
Brillouin-zone was only sampled at the $\Gamma$-point.  Calculations
of the phonon modes and frequencies were performed with
density-functional perturbation theory using the \textsc{Quantum
  ESPRESSO} code \cite{RevModPhys.73.515}.  The electronic orbitals
were expanded in a plane-wave basis set with a kinetic energy cutoff
of 60 Ry.  The Brillouin zone (BZ) integrations in the electronic and
phonon calculations were performed using Monkhorst-Pack meshes. We
refer to meshes of $k$-points for electronic states and meshes of
$q$-points for phonons.  The EPC matrix elements were computed in the
first BZ on a $8\times8\times8$ $q$-mesh ($8\times8\times4$ $q$-mesh)
using individual EPC matrices obtained with a $24\times24\times24$
($24\times24\times16$) $k$-points mesh for fcc (hcp).  Convergence
tests of the EPC calculations were performed with up to
$12\times12\times12$ $q$-meshes for fcc and $8\times8\times8$
$q$-meshes for hcp.

The total phonon zero-point energy was calculated as $E_{ph} =
\frac{1}{2} \sum_i \hbar \omega_i$ over a range of volumes $V$, where
$\hbar \omega_i$ are the individual phonon energies.  We fitted the
results to a polynomial and differentiated with respect to $V$ to
obtain the zero-point enthalpy as $H_{ph} = E_{ph} - V dE_{ph}/dV$.

$T_{c}$ was calculated using the Allen-Dynes
equation,\cite{PhysRevB.12.905}
\begin{equation}
  T_{c}= {\Omega_{log} \over 1.2}
  \exp \left(- \frac{1.04(1+\lambda)}{\lambda - \mu^{*}(1+0.62 \lambda)} \right),
\end{equation}
with the standard value $\mu^{*} = 0.13$.  The normalized weighting
function of the Eliashberg theory \cite{PhysRev.167.331} is
\begin{equation}
  g(\omega) = \frac{2}{\lambda \omega} \alpha^{2} F(\omega).
\end{equation}
The parameter $\lambda$ is a dimensionless measure of the strength of
$\alpha^{2} F$:
\begin{equation}
  \lambda = 2 \int^{\infty}_{0} d \omega\,  \alpha^{2} F(\omega) / \omega,
\label{eqn3}
\end{equation}
and the logarithmic average frequency, $\Omega_{log}$ in units of K,
is
\begin{equation}
  \Omega_{log} = \exp \left( \int^{\infty}_{0} g(\omega) \ln \omega \, d\omega  \right).
\end{equation}

\newpage

\subsection{Crystal structure details}

\textbf{tetragonal PtH at 46.6 GPa}

lattice vectors:

$a$ = $b$ = 3.422 \AA, $c$ = 2.782 \AA\

atomic positions:

\begin{tabular}{l c c c}
Pt & 1/2 & 0 & 3/4 \\
Pt & 0 & 1/2 & 1/4 \\
H & 0 & 0 & 0 \\
H & 1/2 & 1/2 & 1/2 \\
\end{tabular}

\vspace{1cm}

\textbf{fcc PtH at 40.6 GPa}

lattice vector:

$a$ = 2.843 \AA\

atomic positions:

\begin{tabular}{l c c c}
Pt & 1/2 & 1/2 & 1/2 \\
H & 0 & 0 & 0 \\
\end{tabular}

\vspace{1cm}

\textbf{hcp PtH at 48.1 GPa}

lattice vectors:

$a$ = 2.789 \AA, $c$ = 4.733 \AA\

atomic positions:

\begin{tabular}{l c c c}
Pt & 1/3 & 2/3 & 1/4 \\
Pt & 2/3 & 1/3 & 3/4 \\
H & 0 & 0 & 0 \\
H & 0 & 0 & 1/2 \\
\end{tabular}

\newpage

\subsection{Phonon enthalpy contribution}

\begin{figure}[ht]
\centering
\includegraphics[width=0.8\textwidth, angle=0]{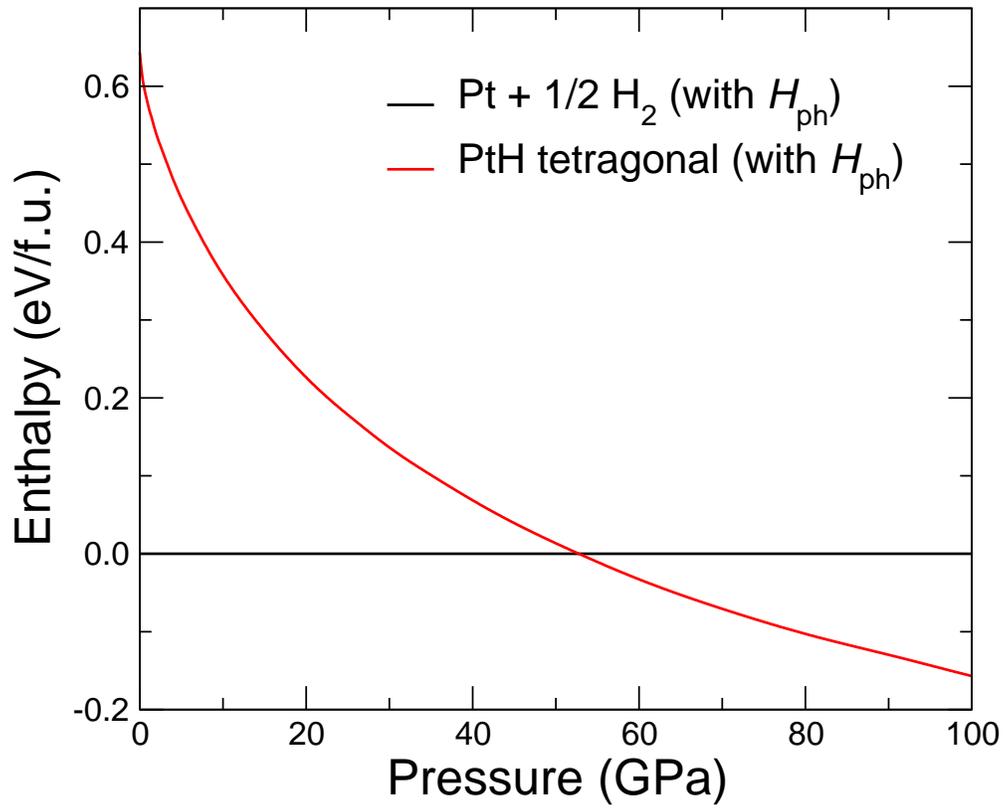}
\caption{Comparison of the enthalpies of Pt + $\frac{1}{2}$ H$_2$ and the
  tetragonal phase of PtH with $H_{ph}$ included.
}
\label{figS1}
\end{figure}

Figure \ref{figS1} shows the effect of the nuclear zero-point enthalpy
(ZPE) $H_{ph}$, as defined in the Methods section, on the formation
pressure of PtH.  The difference in total enthalpy between tetragonal
PtH and Pt + $\frac{1}{2}$ H$_2$ is plotted.  As can be seen, the inclusion of
$H_{ph}$ shifts the formation pressure up to 51.8 GPa, far beyond the
18.7 GPa obtained when the ZPE is neglected.  Such a shift seems
unreasonably high, and we therefore decided to calculate the effect of
$H_{ph}$ for a material with a well-established formation pressure.

\begin{figure}[ht]
\centering
\includegraphics[width=0.7\textwidth, angle=0]{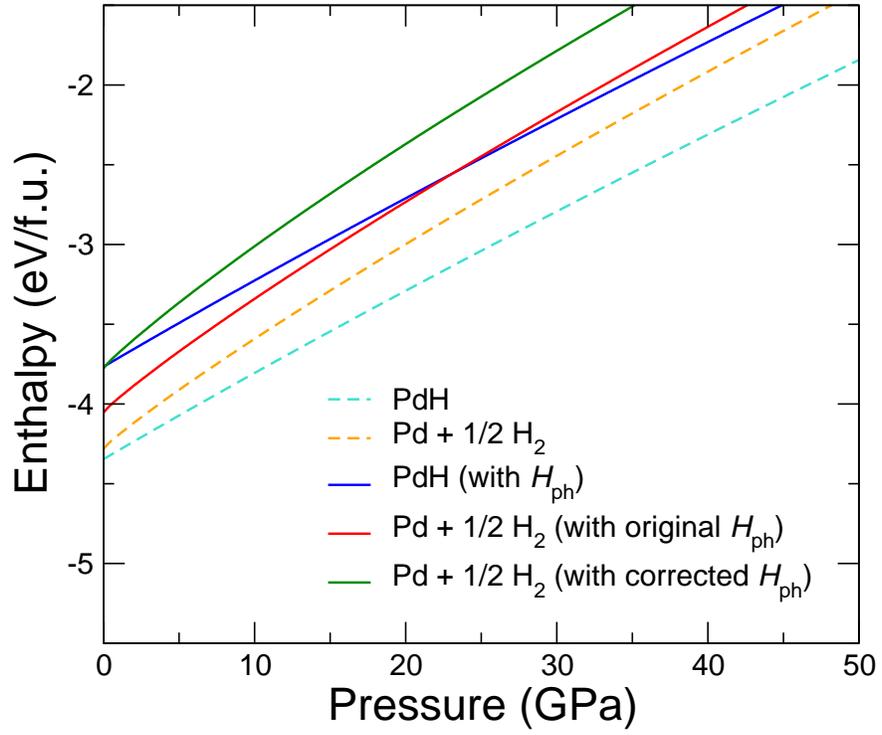}
\caption{Comparison of the enthalpies of Pd + $\frac{1}{2}$ H$_2$ and PdH without
  $H_{ph}$, with $H_{ph}$, and with the corrected $H_{ph}$ contribution.
}
\label{figS2}
\end{figure}

PdH is known to form close to ambient pressure \cite{Stritzker:1972}.
Figure \ref{figS2} shows that the enthalpy comparison of PdH with Pd +
$\frac{1}{2}$ H$_2$ yields a formation pressure in the vicinity of 0 GPa,
actually slightly below it.  However, once $H_{ph}$ is included, the
formation pressure shifts to 23 GPa (Fig.\ \ref{figS2}), in
contradiction with experiment.

\begin{figure}[ht]
\centering
\includegraphics[width=0.9\textwidth, angle=0]{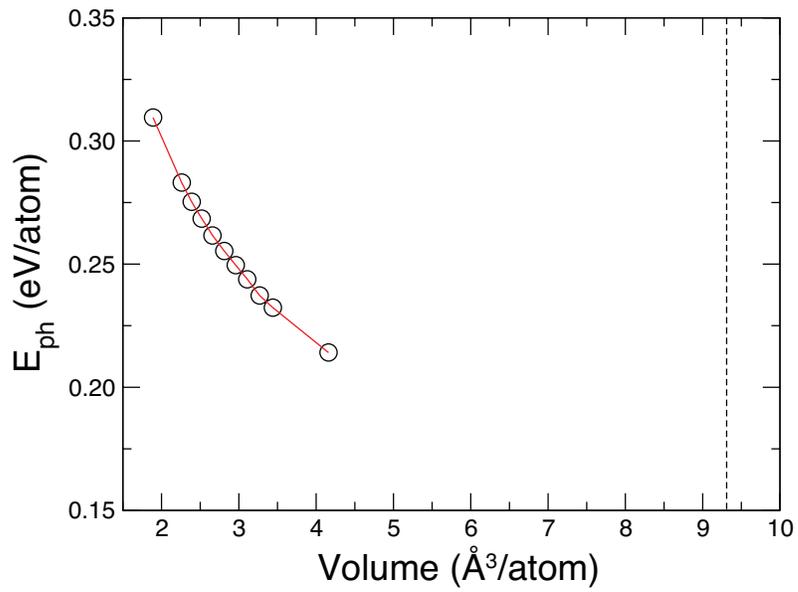}
\caption{$E_{ph}$ for solid H$_2$ is plotted as a function of
  volume. The calculated data are shown as black circles. The connecting
  red lines are merely a guide to the eye. The vertical dashed line
  indicates the equilibrium volume of hydrogen at zero pressure.}
\label{figS3}
\end{figure}

What is the source of this discrepancy? The zero-point enthalpy
$H_{ph}$ contributions from the metal and metal hydride can be rather
precisely determined computationally, in particular because there is
very little uncertainty about the slope $dE_{ph}/dV$, the phonon
pressure term, which is multiplied by the volume and added to the
phonon energy to yield the phonon enthalpy $H_{ph}$.  For Pt and PtH,
the phonon energy $E_{ph}$ is found to depend linearly on the cell
volume, and so $dE_{ph}/dV$ remains virtually constant throughout the
pressure range studied.  The main uncertainty stems from the
contribution to $H_{ph}$ from H$_2$, a system which is notoriously
difficult to describe properly within density functional theory at low
pressures \cite{Ceperley :1987}.  As a consequence, it is hard to
determine the phonon energy $E_{ph}$ accurately at low pressures, and
furthermore, the non-linear dependence of $E_{ph}$ on $V$ in solid
hydrogen (see Figure \ref{figS3}) prevents us from obtaining a
reliable value of $dE_{ph}/dV$ at lower pressure.

To correct for the error in $E_{ph}$ and in the slope $dE_{ph}/dV$, we
therefore determined a correction factor which multiplies $H_{ph}$ for
H$_2$, in order to recover the known formation pressure of PdH.
Figure \ref{figS2} shows the enthalpy when $H_{ph}$ for H$_2$ is
scaled by a factor of 3.3, which yields a reasonable formation
pressure in the vicinity of 0 GPa.

\begin{figure}[ht]
\centering
\includegraphics[width=0.7\textwidth, angle=0]{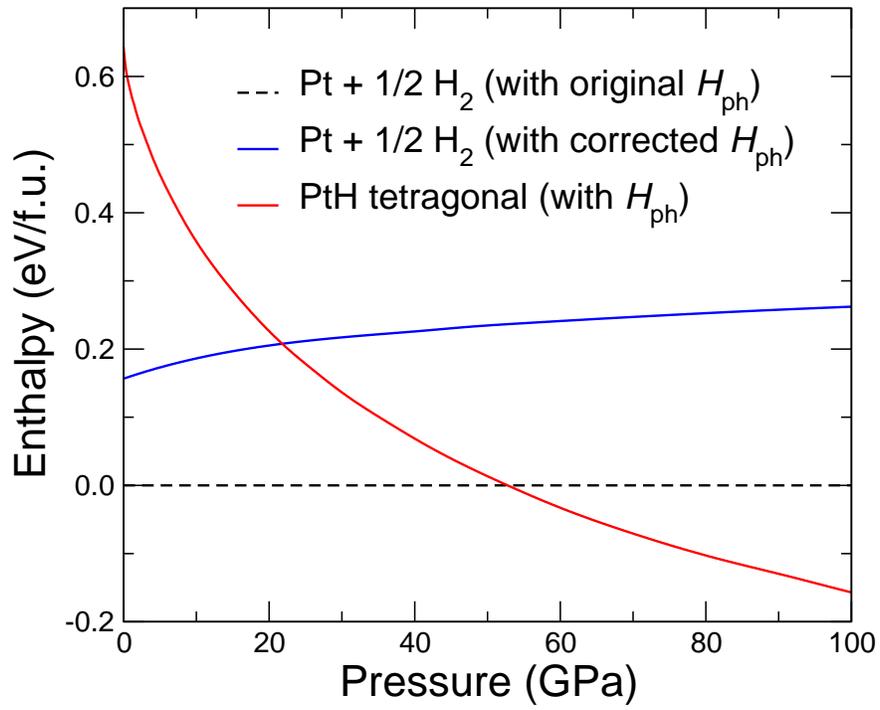}
\caption{Comparison of the enthalpies of Pt + $\frac{1}{2}$ H$_2$ and
  tetragonal-PtH, including the correction discussed in the text.}
\label{figS4}
\end{figure}

Applying the same scaling correction to $H_{ph}$ for H$_2$ in
the enthalpy comparison for PtH (Fig.\ \ref{figS4}), we obtain a
formation pressure of 21.5 GPa, corresponding to a much more
reasonable shift of 2.8 GPa due to the effects of ZPE.

\newpage

\newpage

\newpage

\newpage

\end{document}